\documentclass[preprint,12pt]{elsarticle}

\usepackage[top=2.5cm,bottom=2.5cm,left=2.5cm,right=2cm]{geometry}

\usepackage{graphicx}
\usepackage{amsmath}
\usepackage{colortbl}
\usepackage[dvipsnames,svgnames,xcdraw, table]{xcolor}
\usepackage{longtable,tabularx}
\usepackage{tikz}
\usepackage{scalerel}
\usepackage{subfigure}

\definecolor{pearl}{rgb}{0.94, 0.92, 0.84}

\usepackage{amssymb}

\journal{Journal}

\begin{document}

\begin{frontmatter}



\title{Flow imaging as an alternative to pressure transducers through vision transformers and convolutional neural networks}


\author[inst1]{Renato F. Miotto}

\author[inst1]{William R. Wolf}
\affiliation[inst1]{organization={School of Mechanical Engineering, University of Campinas},
            city={Campinas},
            state={São Paulo},
            country={Brazil}}

\begin{abstract}
In this work, we propose a framework whereby flow imaging data is leveraged to extract relevant information from flowfield visualizations. To this end, a vision transformer (ViT) model is developed to predict the unsteady pressure distribution over an airfoil under dynamic stall from images of the flowfield. The network is capable of identifying relevant flow features present in the images and associate them to the airfoil response. Results demonstrate that the model is effective in interpolating and extrapolating between flow regimes and for different airfoil motions, meaning that ViT-based models may offer a promising alternative for sensors in experimental campaigns and for building robust surrogate models of complex unsteady flows. In addition, we uniquely treat the image semantic segmentation as an image-to-image translation task that infers semantic labels of structures from the input images in a supervised way. Given an input image of the velocity field, the resulting convolutional neural network (CNN) generates synthetic images of any corresponding fluid property of interest. In particular, we convert the velocity field data into pressure in order to subsequently estimate the pressure distribution over the airfoil in a robust manner.
\end{abstract}











\end{frontmatter}


\section{Introduction}
\label{sec:intro}

Pressure is a thermodynamic property that plays a key role in fluid mechanics. It is of utmost importance in aerodynamic load prediction \citep{miottoJFM2022, miottoAIAAJ2022, Pressure_drag_rough_wall}, noise generation \citep{Moreau_Roger_pressure_noise, Azarpeyvand_DSnoise_2022}, flow instability and turbulence \citep{Joshi2014_turbulence, Wu2019_turbulence, Xu2020_instabilities}, and flow control \citep{Deem2020}. The increase of time resolution in velocity field measurements during the last decade has opened the path to obtaining instantaneous pressure fields by combining experimental data with the flow governing equations \citep{van_Oudheusden_2013, vanGent2017}. But often, the available high-speed cameras and particle image velocimetry (PIV) systems do not offer a high enough sampling frequency for unsteady flows of practical interest, which hinders the previous approaches. In case of poor time resolution, the pressure field can be obtained by solving a Poisson equation if the flow is incompressible. But still, the missing time information poses specific constraints in the boundary conditions \citep{VanderKindere2019, CHEN2022110647}.

Recent developments on machine learning have recently been paving the way to new interesting research avenues without the need of time-resolved field measurements to estimate pressure. Using velocity-probe measurements with high sampling frequency, \citet{Jin2020} combined proper orthogonal decomposition (POD) with recurrent neural networks to reconstruct the spatial distribution of velocity, retrieving the time resolution from PIV. \citet{Raissi_etal_2020} used a physics-informed neural network (PINN) \citep{RAISSI2019686} to construct computationally efficient and fully differentiable surrogates for velocity and pressure fields from the transport of passive scalars. Their technique allows the extraction of quantitative information from available flow visualizations such as the transport of dye or smoke in physical systems and contrast agents in biological systems. PINNs were also used to quantify velocity and pressure fields from tomographic background oriented Schlieren \citep{Cai2021} and PIV images \citep{Wang_etal_2022}.

It is not only with regard to the flowfield that pressure is important, but also its distribution on aerodynamic surfaces. For instance, pressure distribution data is crucial for aerodynamic design optimization \citep{MARTINS2022105391} and many surrogate models have recently been developed to reduce the computational time involved in predicting aerodynamic forces during the design process \citep{lui_wolf_2019, HUI2020105949, LEI2021107101, Qiuwan_etal_2022_CNN}. In the context of unsteady aerodynamics, pressure distribution data are used to provide insights on the turbulent structures responsible for the far-field noise generated by unsteady airfoils \citep{Moreau_Roger_pressure_noise, Raus_etal_DS_noise_2022, Azarpeyvand_DSnoise_2022} and also to estimate the instant at which the flow separates near the leading edge \citep{Ramesh_et_al_2014, Deparday_Mulleners_2019, Kay_etal_LESP_2022}. This is particularly useful for developing control strategies for dynamic stall \citep{Chandrasekhara_2007, Sedky_etal_2020}. It is noteworthy that other methodologies that do not depend on pressure to indicate flow separation near the leading edge have been recently proposed (see \citep{Sudharsan_BEF_2022} and \citep{Deparday_LESP_2022} for example), but they are not easily accessible outside of the laboratory environment.

Recently, \citet{HUI2020105949} used a convolutional neural network (CNN) to predict steady pressure distributions over different airfoil profiles. However, the unsteady pressure distribution data constitutes a much more difficult and costly problem because the spatio-temporal complexity of unsteady separated flows renders them difficult to investigate and fully characterize. It comes as no surprise that, thus far, unsteady separated flows have defied obtaining general analytical solutions. During dynamic stall, for example, the generated unsteady aerodynamic loads play a critical role in determining both the mechanical life span and performance of unsteady lifting devices such as wings, helicopter rotors and wind turbine blades \citep{Corke_2015}. To control these loads, it is required both an understanding of the unsteady flow conditions as well as providing mechanisms for prescribing the ensuing blade-flow  interactions. 

Given the importance of quantifying the unsteady forces in engineering systems, the present work aims to provide an alternative to obtain them in experimental settings. Dynamic stall is taken here as an example since the flow has a common structure owning certain complexity as well. Despite that, the concepts applied herein can be easily extended to other branches of fluid mechanics where a regression task is involved. To this end, we chose CNNs and vision transformers (ViTs) to study the flowfields from our high-fidelity numerical simulations of dynamic stall \citep{miottoJFM2022, miottoAIAAJ2022} due to their fewer connections and parameters \citep{Krizhevsky_etal_2012, ViT}. Here, we aim at reusing or transferring information from previously learned tasks to extract quantitative information from available flow visualizations. Different architectures are employed in the hope of finding the mapping relation between the flow structures and the underlying airfoil responses. Based on any fluid property from the unsteady flowfield, the network between the existing flow structures and some concerned flow feature is constructed. The ViT or CNN-based deep learning methods, then, link the map of fluid properties to the aerodynamic loadings, which represents the feature learned from the flowfield. Precisely, we sought a model capable of predicting the distribution of pressure coefficient ($C_p$) over the airfoil. Thus, the present models could serve as an alternative to pressure taps or surface-mounted pressure transducers in acquiring surface pressure data. 

CNNs have been successfully applied to identify features in fluid flows by \citet{Strofer_2019}. \citet{Jin_etal_2018} designed a CNN architecture to predict the velocity field around a cylinder using measurements of the surface pressure as input. \citet{Ye_etal_2020} used the classical simple network LeNet-5 to predict the pressure on a cylinder from the velocity distributions along its wake. CNNs were also shown to be viable alternatives for detecting shock waves, with less time consumption than traditional methods \citep{LIU20191}. This class of network was also employed in a new technique to extract underlying flow features from the original flowfield data, as proposed by \citet{Obayashi_etal_2021}. These authors made use of the nonlinear decomposition from the CNN process to extract flow features different from those of proper orthogonal decomposition in each mode. \citet{Guastoni2021} and \citet{Guemes2021} used CNNs to predict two-dimensional instantaneous velocity-fluctuation fields at different wall-normal locations from wall measurements. The studies mentioned above demonstrate the possibilities that CNNs exhibit for feature detection in fluid mechanics.

In addition to CNNs, the emerging of transformer-based architectures could provide more ideas for leveraging surrogate models in fluid dynamics. This line of research, though, is more scarce in the literature. For instance, among the few studies available, we mention that of \citet{transformer_TBL}, \citet{transformers_forcast} and \citet{transformer_turbulence_grid}. The former authors demonstrated that transformers can efficiently predict the temporal evolution of extremely coarse velocity fields from turbulent boundary layers. Such predictive ability of transformers was also evaluated by \citet{transformers_forcast}. \citet{transformer_turbulence_grid}, in turn, employed a ViT \citep{ViT} along with other techniques to predict turbulence dynamics on coarsened grids. It is still unclear how far the applications of ViTs in fluid mechanics can reach. Above all, more studies need to be carried out to determine their applicability and reliability for fluid flows, not only on turbulence modeling, but also on feature characterization.


The quest to amplify the scope of information extracted in experimental fluid mechanics is the major pursuit of the current work and the aforementioned regression task to predict the unsteady aerodynamic loading is only one step towards this goal. Differently from the cited references, here, we propose a framework whereby numerical simulation data is leveraged to extract relevant information from experimental flowfield 
visualizations. This is somewhat parallel to the recent work by \citet{Wang_PoP_2021, Wang_AIAAJ_2022} that bridges the gap between experiments and numerical simulations through a data fusion approach. However, differently from the previous references, here, we embrace computer vision techniques to this end. In particular, we apply ViTs and CNNs to predict the unsteady pressure distribution over an unsteady airfoil. Despite substantial advances in experimental fluid mechanics, the use of measurements to reliably infer fluid properties, such as density, velocity, pressure or stress fields, is not a straightforward task. This information, on the other hand, comes natural to CFD practitioners. So, in this work, we also address the question of using the information learned from numerical simulation datasets to extract fluid properties from experiments that, until then, would be very complicated or even impossible to obtain. With that, not only we are able to predict surface pressure distributions, but also obtain the entire pressure field or any other property of interest from visualizations of the velocity field.

\section{Methodology}

In this work, we train a deep neural network model to capture relevant flow structures and establish a mapping relationship between these structures and the underlying airfoil response, here expressed in terms of the unsteady pressure coefficient ($C_p$) distribution over the airfoil suction side. Both the location and the morphology of the flow structures with respect to the airfoil must be properly inferred by the neural network for an accurate estimation of the aerodynamic loads. For that, we use ViTs and CNNs for their success in identifying flow features.

In addition to computing a regression model to obtain the aerodynamic loads from flow images, we are also interested in expanding the amount of information that could be extracted from experimental datasets. While these two things might be somewhat related, we refer to them separately. This is because a model is created to obtain the airfoil response, and another one to extract further information from the flow. Here, for example, this other information consists in images. For this, one needs to train a neural network capable of receiving images of the flow as input and generating another image as output. The motivation for this image-to-image translation stems from the fact that any physical properties can be obtained from a numerical simulation, while the amount of information extracted from experiments can be more limited or even too complicated to be acquired. Thus, this model can be trained with some numerical or experimental data for later use in experimental campaigns. In case it is impossible to obtain annotations in an experiment, the model can be trained only with annotated field samples from CFD simulations for a later operation in an experimental setup. 


In what follows, we split the methodology explanation in two parts: one for the regression task related to the prediction of the unsteady aerodynamic loads, and the other for the image synthesis. A PyTorch code implementation with the present methodology is available at {\em link to be added after review}.



\subsection{Regression task}

\subsubsection{Input data}
\label{sec:CNN_input_data}

\begin{figure}[!htb]
     \centering
    \subfigure[$u$-velocity]{
    \includegraphics[width=0.22\textwidth]{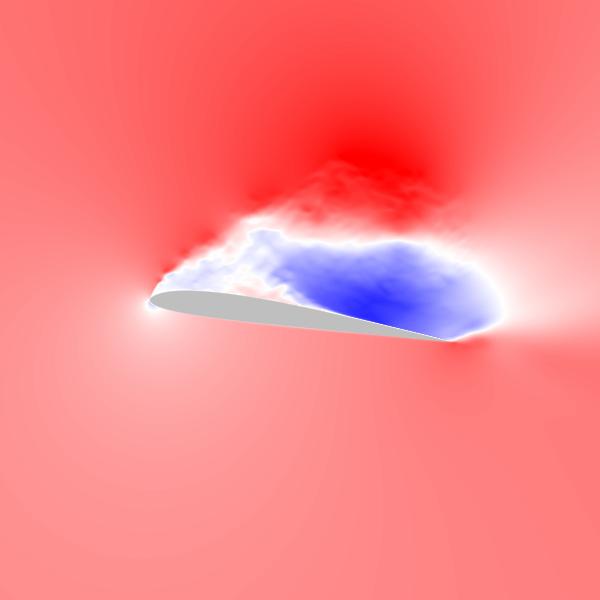}
    }
    \subfigure[$v$-velocity]{
    \includegraphics[width=0.22\textwidth]{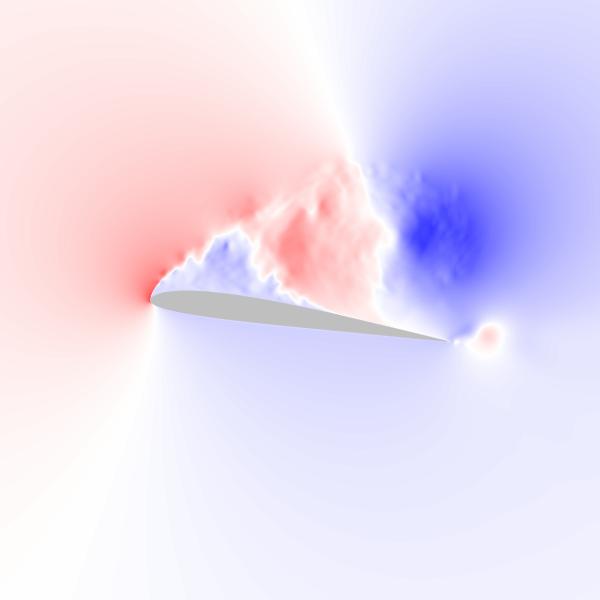}
    }
    \subfigure[$z$-vorticity]{
    \includegraphics[width=0.22\textwidth]{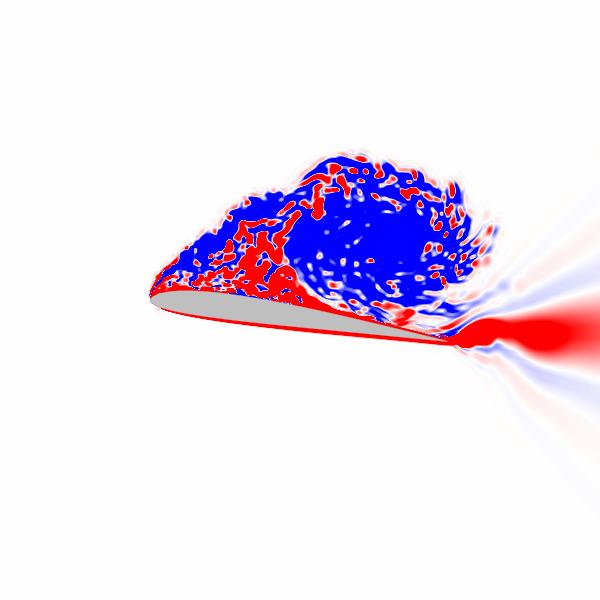}
    }
    \subfigure[$C_p$]{
    \includegraphics[width=0.22\textwidth]{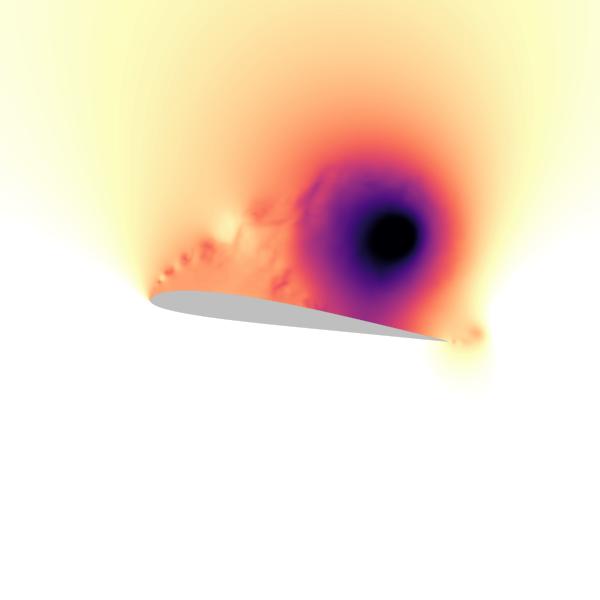}
    }
    \caption{Examples of images used as input to the models. These images correspond to different flow properties for the same instant of a given flow.}
    \label{fig:CNN_input}
\end{figure}

The starting point of the method is the input images of flowfields obtained from 
numerical simulations or experiments, such as those from a 2D cross section.
The images are obtained from the high-fidelity simulations of \citet{miottoJFM2022, miottoAIAAJ2022} in terms of spanwise-averaged flowfields. In these references, the authors were interested in understanding the mechanisms of dynamic stall onset and the conditions for pitch and plunge equivalence to occur. 
Any physical property of interest could be used as input to the network. Here, we train several models
by varying the number of outputs, which is determined by the desired task (further discussion provided in Sec. \ref{sec:architecture}), and the type of fluid property considered as input. One model uses the $u$- and $v$-velocity components as input, while another uses the $z$-vorticity field, and a last one uses the pressure coefficient distribution $C_p$. The rationale for choosing these quantities is that the velocity and vorticity fields can be directly obtained experimentally, through PIV techniques, and the pressure is closely related to the airfoil aerodynamic loads. Figure \ref{fig:CNN_input} shows examples of spanwise-averaged snapshots used as an input to the CNNs and the ViT. 


Table \ref{table:simulations} presents the simulations used to train the neural network, and also employed to assess the capability of the model to interpolate and extrapolate between and beyond simulation parameters. The highlighted rows have no meaning other than to facilitate visualization. In total, five distinct datasets (labeled from DS1 to DS5) are built, all containing 600 $\times$ 600 RGB images of a given physical property. The simulations used for each dataset are identified by the colored circles in Table \ref{table:simulations}. These images consider all simulations of dynamic stall cases reported in Refs. \citep{miottoAIAAJ2022, miottoJFM2022}, which include a periodic plunging airfoil and constant ramp pitching and plunging airfoils for Mach numbers $0.1$ and $0.4$. When generating these images, it is important to keep a fixed range for the contour levels of the property of interest. We used the values $[-2, \ 2]$ for both velocity components, $[-5, \ 5]$ for $z$-vorticity and $[-6, \ 0]$ for $C_p$. The velocity components and length scales are non-dimensionalized by the freestream velocity and chord length.
Finally, this collection of images in each dataset DS was shuffled and arbitrarily divided into groups of cardinality $0.8 \, n$(DS), $0.1 \, n$(DS) and $0.1 \, n$(DS) to form the training, validation, and test sets, respectively.
\newcommand{\circleBlue}[1][blue,fill=blue]{\scalerel*{\tikz \draw[#1] (0,0) circle (4pt);}{\circ}}
\newcommand{\circleCyan}[1][cyan,fill=cyan]{\scalerel*{\tikz \draw[#1] (0,0) circle (4pt);}{\circ}}
\newcommand{\circleTeal}[1][teal,fill=teal]{\scalerel*{\tikz \draw[#1] (0,0) circle (4pt);}{\circ}}
\newcommand{\circleGray}[1][gray,fill=gray]{\scalerel*{\tikz \draw[#1] (0,0) circle (4pt);}{\circ}}
\newcommand{\circleBlack}[1][black,fill=black]{\scalerel*{\tikz \draw[#1] (0,0) circle (4pt);}{\circ}}
\definecolor{whitesmoke}{rgb}{0.96, 0.96, 0.96}
\begin{table}[!htb]
\centering
\begin{footnotesize}
\begin{tabular}{ c c c c c c | c c c c c } 
    \hline
    \multicolumn{6}{c | }{Simulation parameters} & \multicolumn{5}{c}{Datasets} \\
    \hline
    \# & Reynolds & Mach & Motion & Rate / Freq. & No. images & DS1 & DS2 & DS3 & DS4 & DS5 \\
    \hline
    \rowcolor{whitesmoke}
    1  & 60,000 & 0.1 & Plunge periodic & 0.25 & 2049 & \circleBlue & \circleCyan & \circleTeal & \circleGray & \\
    \rowcolor{whitesmoke}
    2  & 60,000 & 0.2 & Plunge periodic & 0.25 & 746 & & & \circleTeal & \circleGray & \\
    \rowcolor{whitesmoke}  
    3  & 60,000 & 0.4 & Plunge periodic & 0.25 & 1024 & \circleBlue & \circleCyan & & \circleGray & \\
    
    4  & 60,000 & 0.1 & Plunge periodic & 0.50 & 1024 & & \circleCyan & & \circleGray & \\
    5  & 60,000 & 0.4 & Plunge periodic & 0.50 & 1024 & & & & \circleGray & \\
    
    \rowcolor{whitesmoke}
    6  & 60,000 & 0.1 & Pitch ramp & 0.05 & 1400 & \circleBlue & \circleCyan &  \circleTeal & & \circleBlack \\
    \rowcolor{whitesmoke}
    7  & 60,000 & 0.1 & Pitch ramp & 0.10 & 2800 & \circleBlue & \circleCyan & \circleTeal & & \circleBlack \\
    \rowcolor{whitesmoke}
    8  & 60,000 & 0.1 & Plunge ramp & 0.05 & 1400 & \circleBlue & \circleCyan & \circleTeal & & \circleBlack \\
    \rowcolor{whitesmoke}
    9  & 60,000 & 0.1 & Plunge ramp & 0.10 & 2800 & \circleBlue & \circleCyan & \circleTeal & & \circleBlack \\
    
    10 & 60,000 & 0.2 & Plunge ramp & 0.05 & 1400 & & & \circleTeal & & \\
    11 & 60,000 & 0.2 & Plunge ramp & 0.10 & 1400 & & & \circleTeal & & \\
    12 & 60,000 & 0.2 & Pitch ramp & 0.10 & 1400 & & & \circleTeal & & \\
    
    \rowcolor{whitesmoke}
    13 & 60,000 & 0.4 & Pitch ramp & 0.05 & 1400 & \circleBlue & \circleCyan & & & \circleBlack \\
    \rowcolor{whitesmoke}
    14 & 60,000 & 0.4 & Pitch ramp & 0.10 & 2800 & \circleBlue & \circleCyan & & & \circleBlack \\
    \rowcolor{whitesmoke}
    15 & 60,000 & 0.4 & Plunge ramp & 0.05 & 1400 & \circleBlue & \circleCyan & & & \circleBlack \\
    \rowcolor{whitesmoke}
    16 & 60,000 & 0.4 & Plunge ramp & 0.10 & 2800 & \circleBlue & \circleCyan & & & \circleBlack \\

    17 & 200,000 & 0.1 & Pitch ramp & 0.05 & 1400 & & & & & \\
    18 & 200,000 & 0.1 & Pitch ramp & 0.10 & 1400 & & & & & \\
    19 & 200,000 & 0.1 & Plunge ramp & 0.05 & 660 & & & & & \\
    20 & 200,000 & 0.1 & Plunge ramp & 0.10 & 2800 & & & & & \\
    \hline
\end{tabular}
\end{footnotesize}
\caption{Databases of high-fidelity simulations of dynamic stall employed in this work and cardinality of the datasets. The colored circles identify the simulations that make up each dataset.}
\label{table:simulations}
\end{table}

Data augmentation is used to artificially increase the size of the training set. Realistic variants of each training instance are generated by shifting, rotating, and resizing every picture through preprocessing layers \citep{Shorten2019}. The transformations applied to the input images are only geometrical and, therefore, preserve the semantics of the images. Moreover, the dynamic stall vortex (DSV) and the entire airfoil are fully framed in all generated instances. 

\subsubsection{Network architectures}
\label{sec:architecture}

In this work we seek for regression models capable of predicting one or more scalar quantities. Particularly, a model is built for predicting the distribution of $C_p$ along the airfoil suction side. The convolutional layers process the flowfield information by extracting features, which are gathered by the fully connected layers to obtain the aerodynamic loadings.
In order to avoid a highly biased model, meaning that it is too simple to learn the underlying structure of the data, we used transfer learning. It is an useful approach to speed up training considerably while also requiring significantly less training data to bootstrap computer vision models \citep{transfer_learning_survey_2019}, and consists of using pre-trained models to leverage features learned on one problem and use them on another problem. Here, we used many architectures pre-trained on the ImageNet dataset, which are readily available for deep learning libraries such as TensorFlow \citep{tensorflow2015, chollet2015keras} and PyTorch \citep{pytorch}. Even though pictures related to fluid flow problems are absent from ImageNet, the method transfers well to the present flowfields involving dynamic stall.

Some CNN architectures are selected to speed up the training process and yield more accurate results in solving the regression problems of interest. Among them, we test the VGG-11-BN \citep{VGG16}, the Inception-V3 \citep{inceptionV3}, the ResNet-50 \citep{ResNet50}, the EfficientNet-B4 and \citep{efficientNet}. For the transformer, we employ the ViT-B/16 \citep{ViT}.
It is important to mention that these networks were trained on classification problems, whereas, here, we are interested in a regression task. Therefore, besides the objective function, it is also necessary to change the top layer of the network to adjust them to our problem.

The present implementation will depend on the number of outputs we want the neural network to have. In this sense, the fully-connected layers are designed to meet our objective of finding the distribution of pressure coefficient over the airfoil suction side. For this, the output needs to be a large array to store the entire load distribution over the airfoil surface. Here, we interpolate the results to 300 points uniformly distributed along the airfoil suction side to keep the data independent of the mesh used in the simulation. Hence, for the pressure distribution task, the output layer is a linear module with 300 units. Finally, the pre-trained models assume that the images are pre-processed in a specific way. Hence, the pre-processing step is not only used for data augmentation, but also to properly scale the pixel range or resize the picture to the size expected by the original model.

\subsubsection{Training procedure}
\label{sec:training}


After instantiating the pre-trained models, the top layer of the fully connected block is replaced by a linear layer that has the desired number of outputs, as mentioned in Sec. \ref{sec:architecture}. One could also have defined an entirely new fully-connected block to add on top of the convolutional base if desired. Then, a feature extraction is performed by keeping the parameters of the model up to this last layer unchanged and training it for 30 epochs. This value is chosen arbitrarily and this step prevents the large gradient updates triggered by the randomly initialized weights from wrecking the learned weights in the convolutional base. 
Also, by doing so, the computational efficiency is improved. In sequence, the network is fine tuned by training all layers for another 170 epochs at a very slow learning rate to make sure that the magnitude of the updates will not wreck the previously learned features. 

Using the framework of maximum likelihood estimation, the mean square error (MSE) cost function is preferred for regression problems. However, here we use the logarithm of the hyperbolic cosine function (logcosh) as it is not strongly affected by occasional wildly incorrect predictions. The NAdam optimizer \citep{Nadam} is also employed with learning rate of $1\times10^{-5}$ accompanied by a scheduler that reduces the learning rate by a factor of 0.6 once learning stagnates for 13 epochs. An early stopping with patience of 50 epochs based on the MSE metric is set to prevent unnecessary computation. In all cases, the batch size is fixed at 28.

\subsection{Image synthesis}

Image synthesis with supervised machine learning is the process of artificially generating images that contain some particular desired content associated with a specific label. The most prominent machine learning model for generating content is known as generative adversarial networks (GANs) \citep{GANs}. The GANs are based on game theory, where two networks compete with each other to generate the best segmentation. One neural network, called the generator, generates new data instances, while the other, the discriminator, evaluates them for authenticity. Differently from the aforementioned networks used for the regression task, GANs are fully convolutional networks (FCN) \citep{FCN}, which are similar to a common convolution neural network. However, the fully connected layers are typically replaced by transposed convolutional layers \cite{deconvolution} which allow one to upsample the input feature map to a desired output feature map using some learnable parameters.

Certainly, the nature of having a discriminator model in a GAN provides output target differences on the pixel level, which emerge from a deeper understanding of the images. However, GANs are often difficult to train and tune \citep{GANs_review}, and a simpler approach to synthesize an image is to use the U-Net \citep{Ronneberger_etal_2015}, a network developed to work with fewer training images and produce accurate biomedical image segmentation. This network resembles an encoder-decoder structure, but with the addition of skip connections that are used to transfer fine-grained information from the low-level layers of the contracting path to the high-level layers of the expanding path. For its simplicity, it is the architecture used in the present work. Usually, the U-Net combines a pixel-wise Softmax over the final feature map with the cross entropy loss function. However, the Softmax is removed in the present framework and the MSE loss function is employed instead.


The velocity field is a good candidate for an input since it can be directly obtained from experiments using PIV. So, we build a U-Net model that takes images of $u$- and $v$-velocity components, concatenated channelwise. The outputs can be any fluid property of interest. Here, we are interested in the spatial distribution of pressure coefficient, $C_p$, as output for its importance in engineering systems, as discussed in Sec. \ref{sec:intro}. Both input and output images have resolution of $256 \times 256$. Similarly to what is done with regression models, data augmentation is employed to artificially increase the number of training samples. This augmentation only applies translation, horizontal flipping and scaling operations. In all training instances, the airfoil and dynamic stall vortex are fully framed. To investigate the interpolation and extrapolation capabilities of the model, the same datasets from Table \ref{table:simulations} are used to train the network. 

The U-Net is trained for 100 epochs using the NAdam optimizer \citep{Nadam} with a learning rate of $3\times10^{-4}$ and a batch size of 8 to minimize the MSE objective. Similar to what was mentioned before, a scheduler that reduces the learning rate by a factor of 0.6 once learning stagnates for 13 epochs is also employed and an early stopping with patience of 50 epochs based on the MSE metric is set to prevent unnecessary computation.

\section{Results}

\begin{figure}[!htb]
    \centering
    \includegraphics[width=\textwidth]{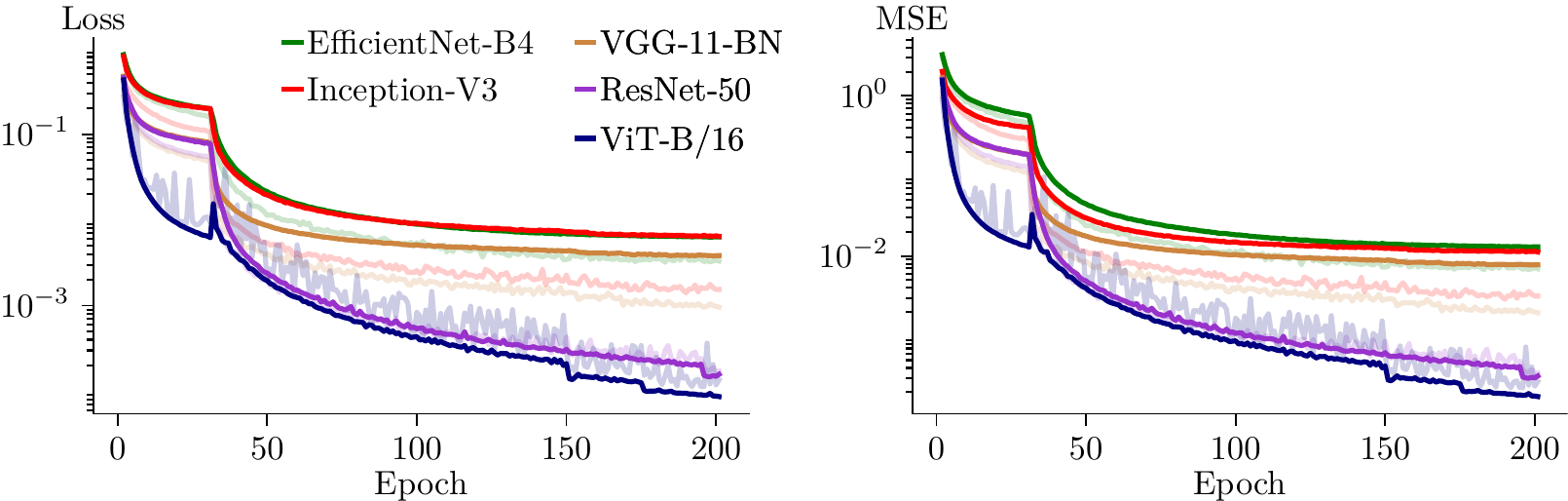}
    \caption{Evaluation of the logcosh loss (left) and mean squared error (right) for all networks on the training (opaque lines) and validation (semi-transparent lines) sets of \circleBlue \space DS1 using images of $u-$ and $v$-velocity components as input to predict the airfoil $C_p$ distribution.}
    \label{fig:training_Cp_distribution}
\end{figure}

In this section, we present the results obtained with our CNN models and the ViT, starting with the performance of each model on \circleBlue \space DS1 (see Table \ref{table:simulations}) training and validation sets as shown in Fig. \ref{fig:training_Cp_distribution}. In this figure, the opaque lines represent the training set, while the semi-transparent lines represent the validation set. At epoch 30, a discontinuity appears in the plots, corresponding to the moment in which fine tuning starts (see Sec. \ref{sec:training} for details). After training, it is evident that the ViT-B/16 network performs better than the other architectures for both training and validation sets. EfficientNet-B4, in turn, is at the opposite extreme. Note that the loss and MSE of Inception-V3 and VGG-11-BN are lower in validation than in training. This is expected and occurs due to the presence of either dropout or regularization in the networks, which inflates their training loss. 
As ViT-B/16 presented a better performance in this task, all the following results will be shown only for this network. Notice that the ViT-B/16 and the ResNet-50 could be trained even further to improve their results, but we decided to keep the training limited to 200 epochs for simplicity, and also considering that satisfactory results were obtained. Training duration varies according to the architecture and the size of the dataset; as an example, for \circleBlue \space DS1, it takes from 9 to 12 hours in a NVIDIA Tesla A100 for each network. 

\begin{figure}[!htb]
    \centering
    \includegraphics[]{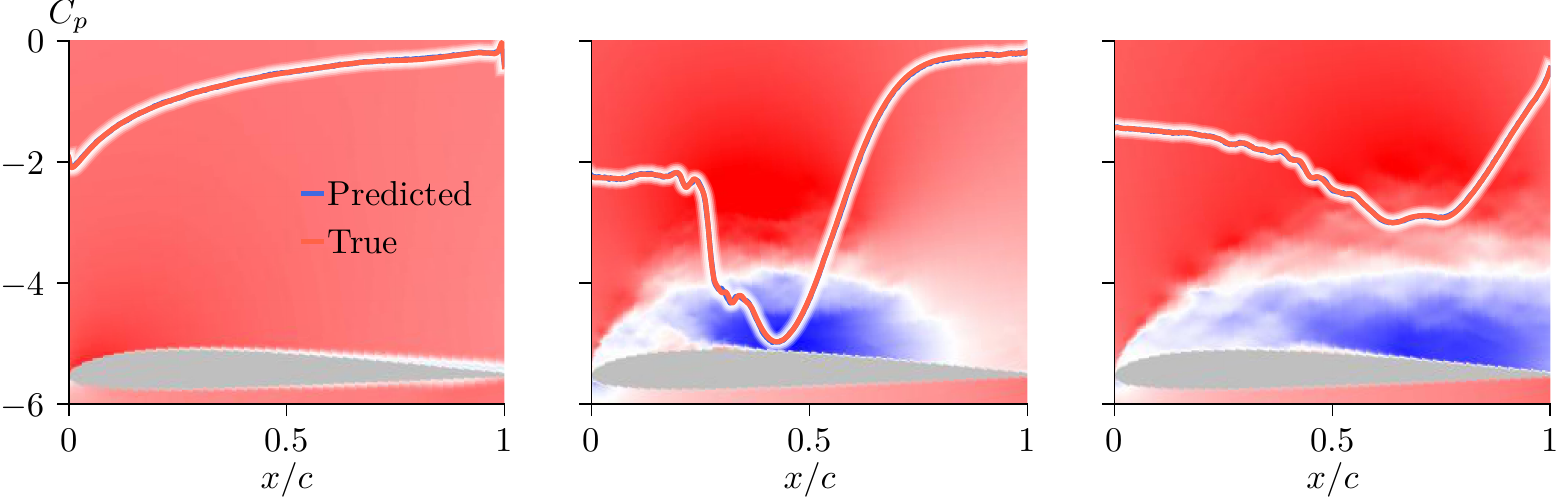}
    \caption{Pressure coefficient distribution over the airfoil suction side and corresponding snapshot of $u$-velocity field (in background) of arbitrarily chosen instances from the \circleBlue \space DS1 test set, never seen by the network.}
    \label{fig:Cp_distribution_testset}
\end{figure}

Results for the $C_p$ distribution over the airfoil suction side for some selected snapshots belonging to the \circleBlue \space DS1 test set are compared against their true values in Fig. \ref{fig:Cp_distribution_testset}.  The model used $u$- and $v$-velocity components as input, but the images reproduced in the graphs only refer to the corresponding $u$-velocity field. The reason behind using velocity as input is because it can be obtained directly from PIV measurements. An excellent agreement between the actual computed distribution of $C_p$ and that predicted by the network is observed, showing that the ViT is capable of processing the pressure information by extracting relevant features from the input image. This information was then translated into surface measurements. A similar conclusion holds for the CNNs (not shown here). The significance of these results is much more optimistic: they show that, if the model generalizes to other flow or motion parameters, the pressure sensors can be partially or totally replaced by a ViT or CNN-based model. This would have a big impact on experimental campaigns because this type of measurement requires the installation of several probes, which can be complicated to assemble or cost prohibitive. Thus, if access to numerical results are provided from similar flows of interest, such data can be used to train models for later use in experimental setups, replacing the need for pressure transducers. In the absence of numerical data, in turn, experimental measurements need to be obtained to train the models. However, due to its ability to generalize to parameter variations, the underlying training process would only require a few experimental runs and, once trained, a model can be reused in future studies. In this case, the transducers can be partially replaced by the  models.

\subsection{Generalization}

\begin{figure}[!htb]
    \centering
    \includegraphics[width=\textwidth]{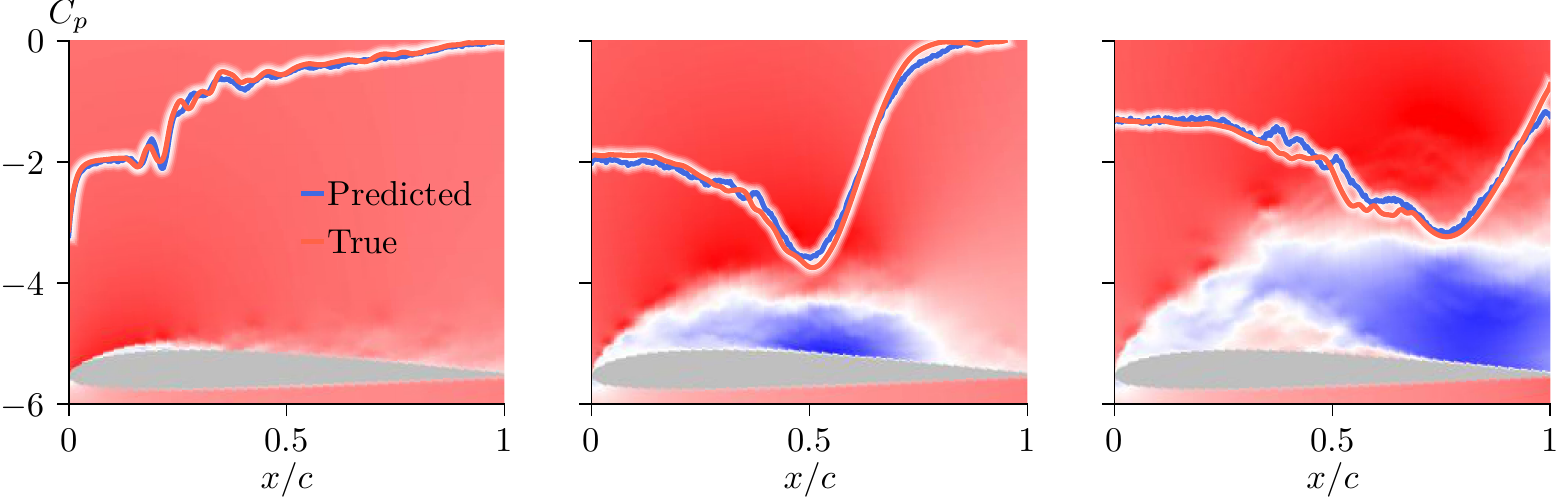}
    \caption{Ability to interpolate between different compressible regimes. The model is trained on the \circleBlue \space DS1 set ($M_{\infty} = 0.1$ and $0.4$) using $u$- and $v$-velocity components as input, and evaluates the surface $C_p$ distribution for arbitrary snapshots from simulation \#10 with $M_{\infty} = 0.2$.}
    \label{fig:interpolate_mach_distribution_uv}
\end{figure}

The previous arguments hold based on the assumption that the model is able to generalize to the flow and motion parameters. Obviously, this hypothesis needs to be verified. 
To answer this question, we begin by investigating the model's ability to predict the pressure distribution in a flow with an intermediate Mach number. We train the model on the \circleBlue \space DS1 dataset, which contains flow snapshots for $M_{\infty} = 0.1$ and $M_{\infty} = 0.4$. Then, a test is performed using the snapshots from simulation \#10, which consists of an SD7003 airfoil under the same pitch-up maneuver reported by \citet{miottoAIAAJ2022}, but at freestream Mach number $M_{\infty} = 0.2$. Despite never seen an image of the dynamic stall problem for this Mach number, the network is capable of predicting the unsteady surface pressure distribution, as shown in Fig. \ref{fig:interpolate_mach_distribution_uv}. This figure shows results for three selected snapshots from simulation \#10. The $u$-velocity field is shown in the background, and the instantaneous surface $C_p$ distributions are presented for the CFD calculation and the network model. 
One can observe that the model predicts the main trends of the pressure distribution with a good accuracy and, hence, it is able to interpolate between different compressible regimes. 
%

The model capability to extrapolate between different airfoil kinematics and flow parameters is also sought. However, this is a more challenging task because depending on the extrapolation parameters, the flow features can be very different from what the network was trained for. For example, if the Mach number is significantly increased, the mechanism of dynamic stall formation will involve shock waves \citep{Bowles_etal_2012, Corke_2015} which have never been seen by the neural network. Hence, the model is not expected to be able to correctly interpret images with semantics very different from those on which it was trained. Difficulties in extrapolating the source domain were also observed by \citet{Erichson_etal_2020_extrapolation}, despite these authors employing a shallow architecture to solve a different task from ours. Later in this work (Sec. \ref{sec:sources_of_variability}), we will explore this subject considering different combinations of simulations in the construction of the training set, through which it is possible to see how the model generalization is impacted when some semantics are not present during training.

\begin{figure}[!htb]
    \centering
    \includegraphics[width=\textwidth]{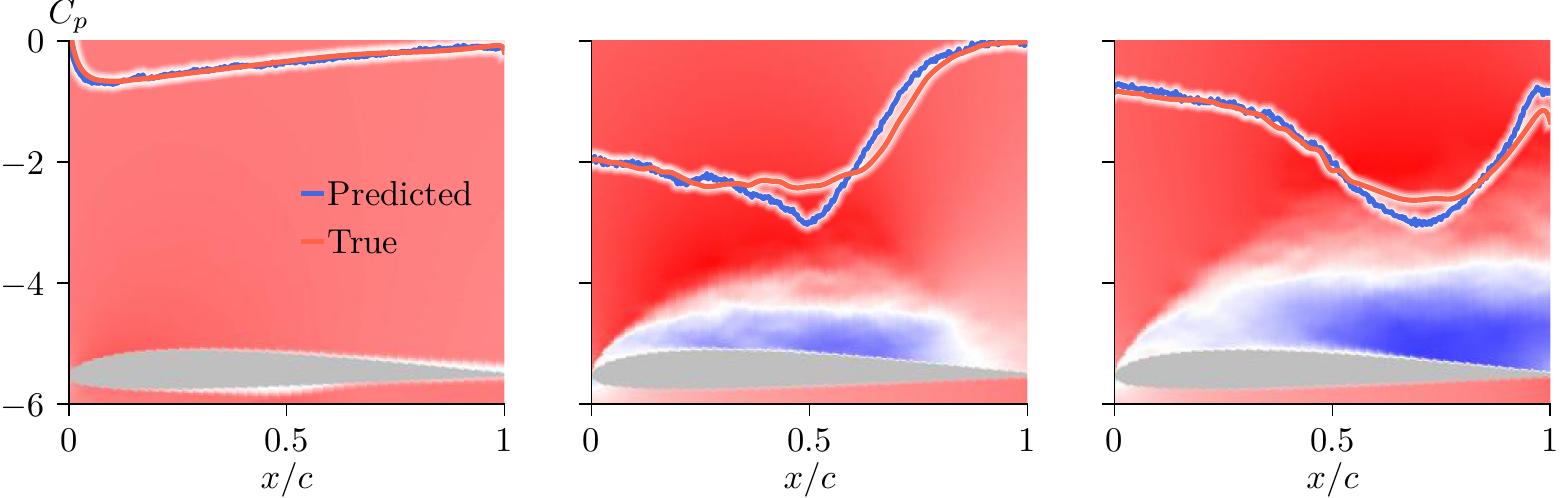}
    \caption{Ability to extrapolate for different compressible regimes. The model is trained with the dataset \circleTeal \space DS3 ($M_{\infty} = 0.1$ and $0.2$) using $u$- and $v$-velocity components as input, and evaluates the surface $C_p$ distribution for arbitrary snapshots for simulation \#3 with $M_{\infty} = 0.4$.}
    \label{fig:extrapolate_mach_uv}
\end{figure}

It was shown that the ViT model can interpolate between different compressible flows for Mach numbers 0.1 and 0.4. Now, the capability of the network is tested for extrapolation of different Mach numbers using the dataset \circleTeal \space DS3 (see Table \ref{table:simulations}) to train the model. This dataset  consists of simulations with $M_{\infty} = 0.1$ and $0.2$, and the extrapolation is performed for the $M_{\infty} = 0.4$ flow from simulation \#3. Results are shown in Fig. \ref{fig:extrapolate_mach_uv}, where it can be seen that, unlike the previous interpolation results, the deviations of the predictions from the true values are larger for the extrapolation task, especially during the transport of the dynamic stall vortex. However, the main trends are captured by the ViT prediction.
%

\begin{figure}[!htb]
    \centering
    \includegraphics[width=\textwidth]{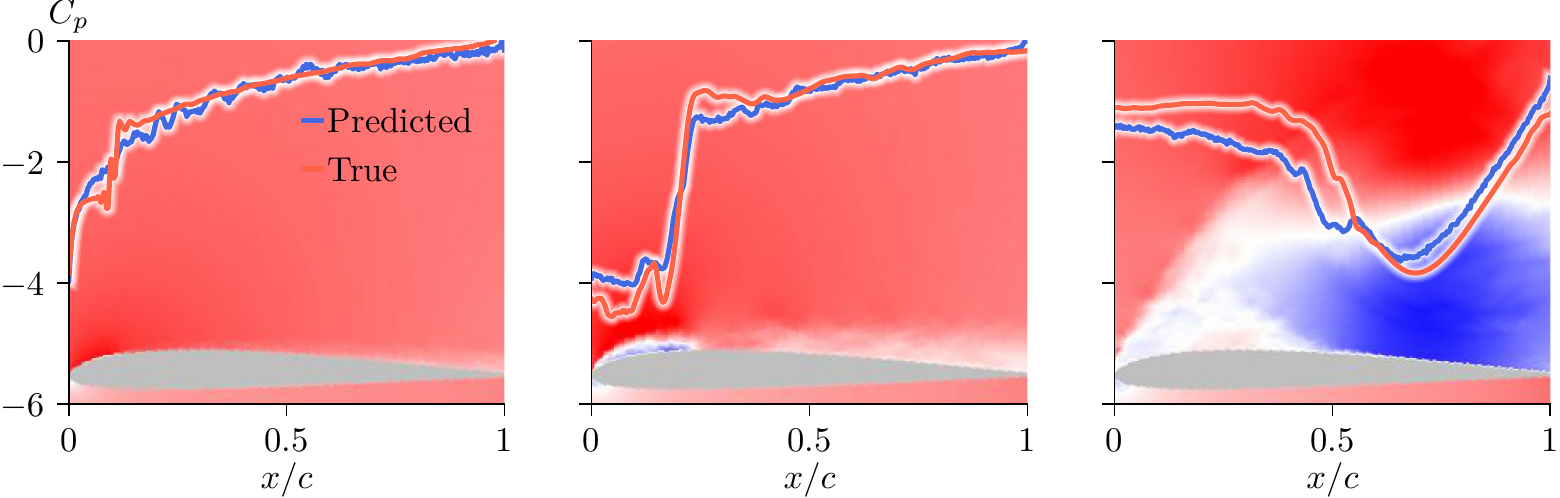}
    \caption{Ability to extrapolate for different Reynolds numbers. The model is trained on the \circleBlue \space DS1 ($Re = 60,000$) using $u$- and $v$-velocity components as input, and evaluates the surface $C_p$ distribution for arbitrary snapshots of simulation \#17 with $Re = 200,000$.}
    \label{fig:extrapolate_reynolds_uv}
\end{figure}

The extrapolation capability of the model is also tested for a different Reynolds number. Figure \ref{fig:extrapolate_reynolds_uv} shows the results of the network trained on the \circleBlue \space DS1 dataset with Reynolds number $Re = 60,000$. The model is tested with simulation \#17 with $Re = 200,000$. Overall, the predicted pressure load distributions follow the same trends as the computed CFD curves. 
In the leftmost plot of figure, a step-like pattern appears in the true pressure distribution for $0.1 < x/c < 0.2$, and this feature is smoothed out by the network. This pressure jump pattern is due fine-scale instabilities that take place in a separation bubble in the vicinity of the leading edge. Such phenomenon was studied in detail by \citet{Benton_Visbal_2018a, Benton_Visbal_2019} and it is also present in the ramp motion simulations at $Re = 60,000$ \citep{miottoAIAAJ2022}. However, as the Reynolds number increases to $Re = 200,000$, the separation bubble becomes smaller \citep{Benton_Visbal_2019} and the resolution of the input image, which has size $224 \times 224$, is not sufficient to represent the fine scales, causing the predicted labels to being smoothed out
\footnote{Although each network expects a specific image size, there is room for improvement in capturing fine-grained scales by zooming into the region of interest. In the results shown here, the airfoil did not occupy the entire width of the image, as shown in Fig. \ref{fig:CNN_input}. In fact, the background images in the results were enlarged from the original, non-resized, $600 \times 600$ input image (Fig. \ref{fig:CNN_input}) to aid in the interpretation of the plots. So, these are not the actual network inputs.}.
%

\subsection{Using different fluid properties as input}
\label{sec:different_property_as_input}

In the previous section, we verified that the trained model is able to interpolate and extrapolate for different flow parameters using velocity field images as input. However, it is observed that the extrapolated results have some discrepancies from the true $C_p$ solutions, especially in the more advanced stages of the dynamic stall vortex. This could be due to a high-variance model or an irreducible error in the velocity data itself. By irreducible data error, we are not saying that the velocity field is incorrect, but rather that the images are inherently too noisy for the regression task. To check whether overfitting occurs due to model complexity, we compared the predictions of all architectures mentioned in Section \ref{sec:architecture} using velocity components as input and found very similar trends (not shown here for brevity). Thus, the possibility of the low extrapolation capacity being related to the complexity of the model was ruled out. This suggests that using another fluid property as input to the network can lead to better generalization.

As it will be shown below, the ViT inaccuracy in extrapolating the pressure distribution is compensated by directly using the pressure field as input. We also test the $z$-vorticity field, but the obtained results had a worse agreement (not shown here) due to the fact that the vorticity field has a wide range of magnitudes throughout the entire domain. For example, near the wall, the non-slip boundary condition implies a region of high vorticity that causes saturation of image levels. This poses a difficulty in model inference for cases that go beyond the source domain ({\em i.e.}, the training data), as the saturation close to the wall does not allow an accurate estimate of the vorticity value in that region. Precisely, our learned task now faces different conditional distributions between the source and target domains \citep{survey_transfer_learning2010}. Although the boundary levels could be increased to avoid their saturation near the wall, relevant information about the separated flow would be lost, which is not a positive trade-off.

\begin{figure}[!htb]
    \centering
    \includegraphics[width=\textwidth]{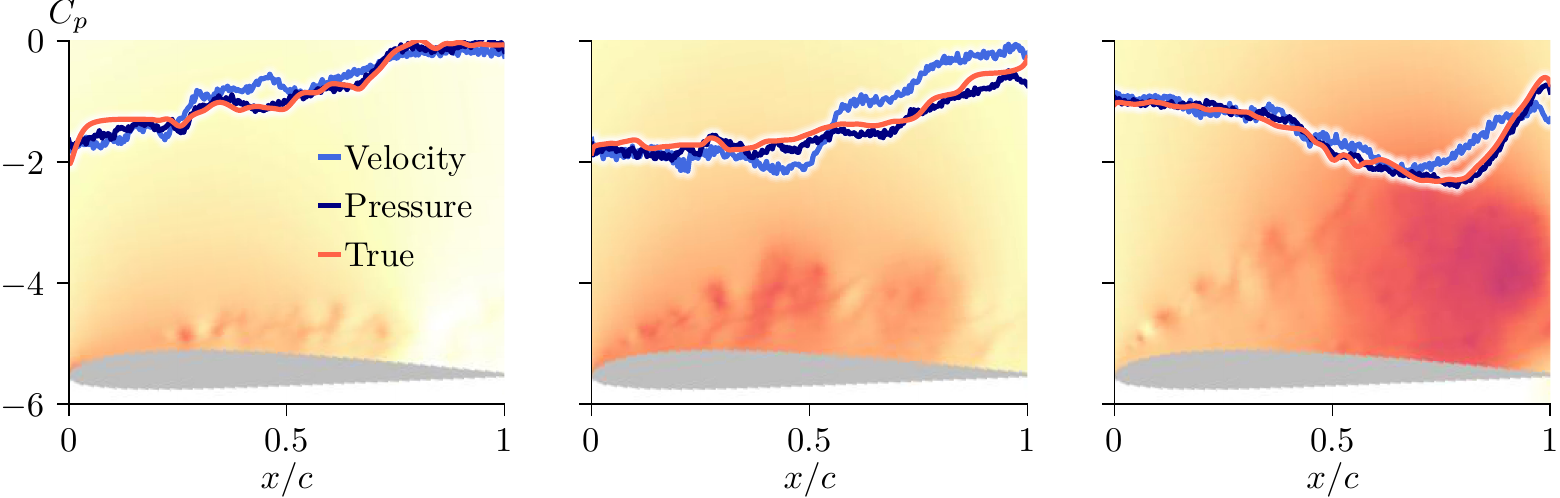}
    \caption{Ability to extrapolate for different compressible regimes. The model is trained with \circleTeal \space DS3 ($M_{\infty} = 0.1$ and $0.2$), and evaluates the surface $C_p$ distribution for arbitrary snapshots of simulation \#13, with $M_{\infty} = 0.4$.}
    \label{fig:extrapolate_mach_uv_vs_pressure}
\end{figure}

\begin{figure}[!htb]
    \centering
    \includegraphics[width=\textwidth]{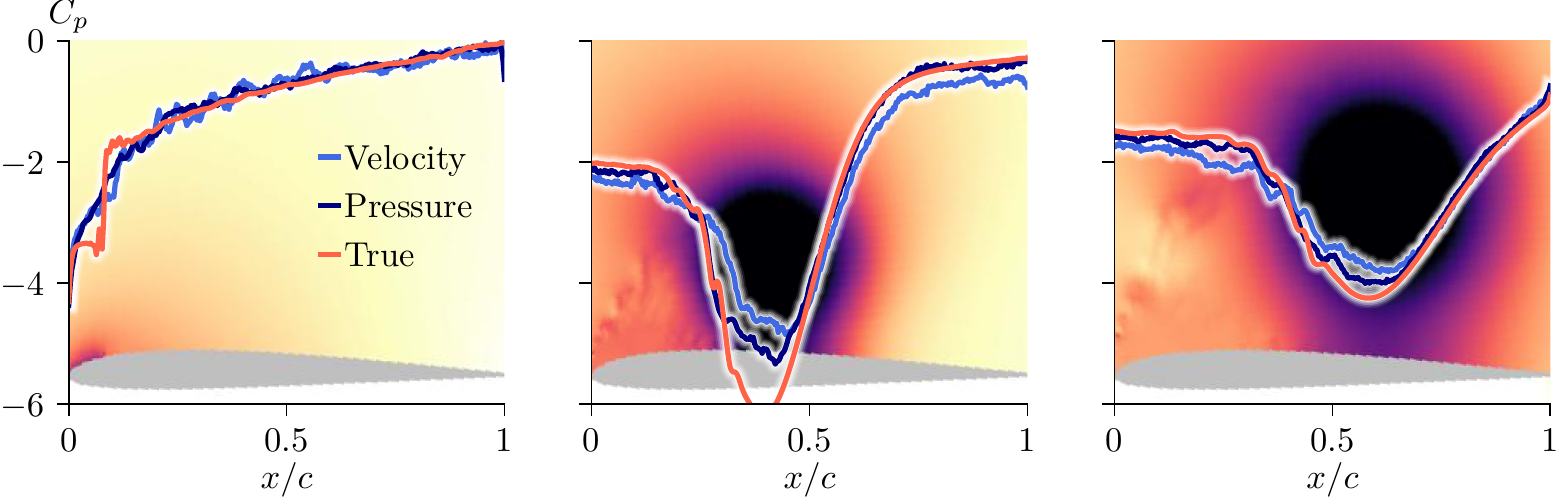}
    \caption{Ability to extrapolate for different Reynolds numbers. The model is trained with \circleBlue \space DS1 ($Re= 60,000$), and evaluates the surface $C_p$ distribution for arbitrary snapshots of simulation \#18 with $Re = 200,000$.}
    \label{fig:extrapolate_reynolds_uv_vs_pressure}
\end{figure}

Figures \ref{fig:extrapolate_mach_uv_vs_pressure} and \ref{fig:extrapolate_reynolds_uv_vs_pressure} compare the results obtained by two models, one trained with velocity components, and  another trained with pressure. Comparisons are provided with the true flow values for extrapolations in terms of Mach and Reynolds numbers. For the Mach number extrapolation (Fig. \ref{fig:extrapolate_mach_uv_vs_pressure}), the dataset \circleTeal \space DS3 ($M_{\infty} = 0.1$ and $0.2$) is used for training the models and we use simulation \#13 as a target with $Ma_\infty = 0.4$. The model employed in the Reynolds number extrapolation is trained with \circleBlue \space DS1 ($Re = 60,000$) and tested with simulation \#18 which is performed for $Re = 200,000$. As can be seen from these figures, predicting the surface distribution of $C_p$ when using the pressure coefficient field as input guarantees a better agreement with the expected results. In the present figures, one can see that the results of the Mach number extrapolation show excellent agreement for all images, and those obtained for the Reynolds number extrapolation are also improved when the pressure field is used as an input instead of the velocity components.

From Fig. \ref{fig:extrapolate_reynolds_uv_vs_pressure}, we still see that the laminar separation bubble is not captured (leftmost figure), but this is expected as the image resolution is the same as before. 
In the second plot of the same figure, the low pressure region of the vortex core is close to the airfoil surface, and the saturation of the image makes it impossible to correctly predict the load along this region. This is a concept shift problem \citep{lemberger2020primer, UDA-Review} analogous to the one encountered when using a model based on z-vorticity, as discussed above. However, unlike what happens with the vorticity field, the low pressure vortex core that leads to the image saturation does not remain close to the wall at all instants of time, which makes the problem simpler to solve. One could, for example, increase the range of the color scale and retrain the model for the new values. It would also be possible to create a system that accepts the maximum and minimum limits of the scale as an input metadata, making it robust to images with different color scales. This metadata could be informed to the fully-connected layers of the existing architecture, or else a hypernetwork could be created using this metadata to return the parameters of the fully-connected layers. 

\begin{figure}[!htb]
    \centering
    \includegraphics[width=\textwidth]{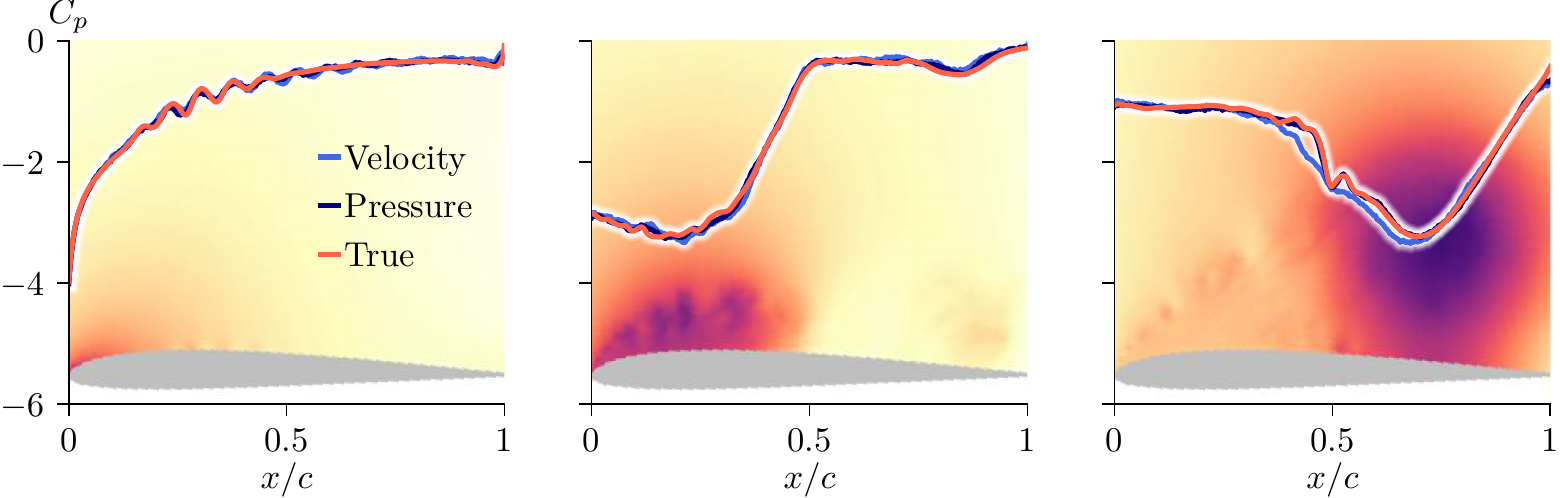}
    \caption{Ability to interpolate between different compressible regimes. The model is trained on \circleBlue \space DS1 set ($M_{\infty} = 0.1$ and $0.4$), and evaluates the $C_p$ distribution for arbitrary snapshots from simulation \#2 with $M_{\infty} = 0.2$.
    }
    \label{fig:interpolate_mach_distribution_uv_vs_pressure}
\end{figure}

Although the scale range information is an interesting solution to the color saturation problem, its implementation is beyond the scope of this work. Here, we want to keep the model as simple as possible as it is a proof of concept. Indeed, in the absence of saturated regions near the airfoil surface, the present ViT-based model proves to be an interesting alternative to physical measurement devices. As the results demonstrate, the model is robust with respect to changes in Mach and Reynolds numbers, and this stems from the fact that, despite the flow being different from those with which the network was trained, there is a common high-level semantics between them. Furthermore, we see that when the $C_p$ field is used as input, the extrapolation capacity of the system improves significantly. The same also occurs with interpolation, as we can see in Fig. \ref{fig:interpolate_mach_distribution_uv_vs_pressure}, where an excellent agreement with the expected result is observed. 

As the results suggest, the success of our predictor depends on its capacity to learn intermediate concepts between raw input and label that are general enough to make sense across a wide range of different but related domains. The intuition is that, assuming there are latent variables that correspond to the true explanatory factors of the observed data, answering questions and learning dependencies in the space of these latent variables is likely to be easier than answering questions about the raw input. In fact, past studies reveal that deep networks learn more transferable representations that disentangle the explanatory factors of variations behind data \citep{bengio2013representation, oquab2014learning}. This helps explain why our deep model is more successful at extrapolating results when compared to the shallow architecture of \citet{Erichson_etal_2020_extrapolation}. However, deep representations can only reduce the cross-domain distribution discrepancy, but not remove it \citep{yosinski2014transferable}. Therefore, recent research on deep domain adaptation further embeds adaptation modules in deep networks that explicitly minimize a distribution discrepancy measure, or use adversarial training to align source and target domains in the representation space \citep{survey_UDA_2020, UDA-Review}.

In the present work, we choose not to use any technique for distribution matching because this involves challenges that are beyond the objectives prescribed here. In fact, domain adaptation methods for high-dimensional regression are still in an early stage and lack an effective approach. For instance, typical domain adaptation methods developed for classification are based on the clustering assumption to get confident decision boundaries and under-perform in regression tasks \citep{chidlovskii2021universal}. One reason for this lies in the fact that regression performances are not robust to feature scaling that occurs in domain adaptation classification methods \citep{chen2021representation}. Thus, applying an improper domain adaptation technique may be ineffective or even harmful in our regression setting.

To overcome the absence of a more principled approach to match the feature distribution across domains, here we leverage the data itself, using multiple simulations, to learn representations that separate the various explanatory sources. Doing so should yield a significantly more robust representation for the complex and richly structured variations that exist in dynamic stall cases. What is shown next is the importance of including possible semantic variations in the training data for a good generalization. Later, we will also discuss alternatives for using this tool in experimental settings where the pressure field cannot be easily obtained to serve as an input to the network.

\subsection{Including sources of variability}
\label{sec:sources_of_variability}

Knowing that pressure is a much more robust property for model generalization, we will use it from now on to show how CNN and ViT predictions are impacted by adding or removing some flow feature from the training set. For instance, any rescaling of the color map caused by the intensity of the dynamic stall vortex, including changes in its topology and location, are known nuisance factors and should be learned at the outset.
This is especially important since the continuous interpretation of an image changes according to these factors, and networks do not transfer well across different feature and label distributions \citep{pedestrians2009, chapter_transfer_learning, survey_transfer_learning2010, UDA-Invariant-representation-2019}. Fortunately, the potential nuisance variability is usually already known in non-stationary aerodynamic problems and can be dealt with early on, without the need to learn it through complex adversarial training \citep{domain_adversarial_nets2016, UDA-Review}. In fact, we know that the flow and the underlying aerodynamic response are influenced by the effects of compressibility \citep{Bowles_etal_2012, Corke_2015, miottoJFM2022} and Reynolds number \citep{Schreck_etal_2002}, as well as the airfoil profile \citep{Sharma_visbal_2019} and its kinematics \citep{Schreck_etal_2002, miottoAIAAJ2022}. Thus, including these sources of variability in the training set improves generalization. 



What we just mentioned above might lead one to believe that any nuances would need to be included in the training set, making the learning process not practical for the task at hand. But this is not entirely true. Note that, although we never varied the Reynolds number during training, the ViT prediction was very accurate in extrapolating this parameter. This stems from the fact that the marginal distribution (distribution of features) between the source and target domains are very similar. At colloquial level, this means that the neural network has learned flow features that are still present in the target domain (Reynolds 200,000 flow, in this case). So, let $X$ and $Y$ denote the (flow) features and target (aerodynamic response), respectively, forming a causal system in which $X$ is the cause for $Y$. To reduce domain discrepancy, we seek for invariant components $\mathcal{T}(X)$ that have a similar probability distribution $P(\mathcal{T}(X))$ across domains. However, it is not clear if the conditional probability $P(Y|\mathcal{T}(X))$ in different domains is also similar when $P(Y|X)$ changes \citep{gong2016domain}. In other words, it is not possible to guarantee that the learnt representation $\mathcal{T}(X)$ will have any relevant information for predicting $Y$ in the target domain. In fact, learning invariant representations alone is not sufficient to guarantee good generalization on the target domain, even after minimizing the source error \citep{UDA-Invariant-representation-2019, survey_UDA_2020}. It is also necessary to align the label distributions, which without target labels is still an open question.

\begin{figure}[!htb]
    \centering
    \includegraphics[width=\textwidth]{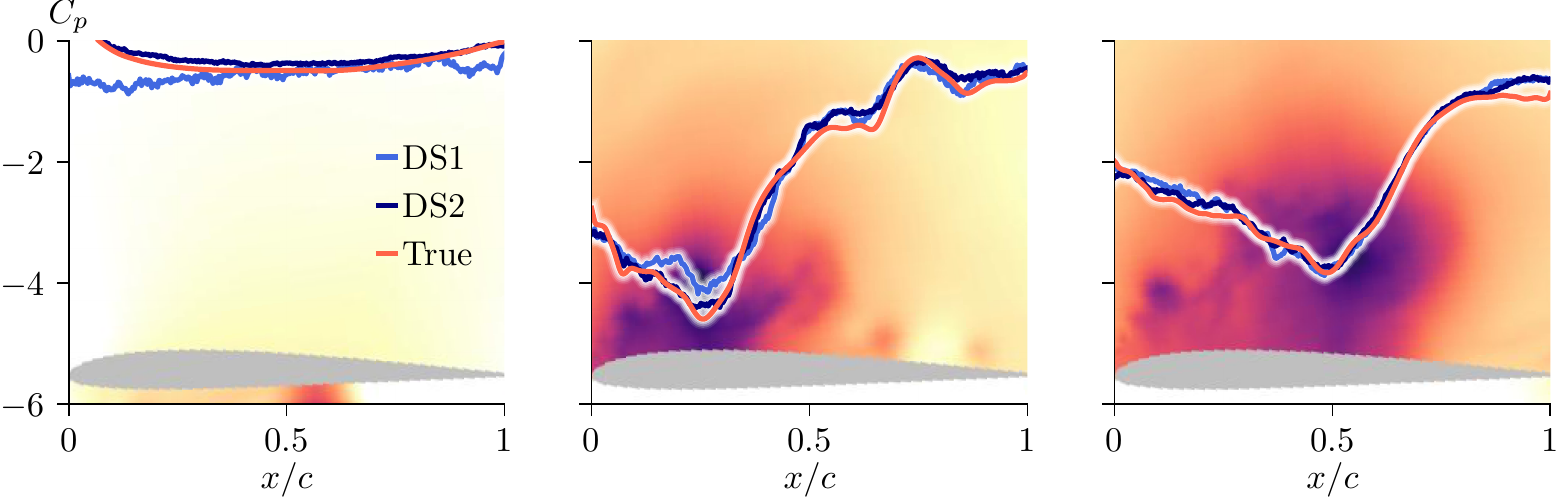}
    \caption{Effects of varying reduced frequency. The model is trained on \circleBlue \space DS1 (reduced frequency 0.25) or \circleCyan \space DS2 (adding a simulation with reduced frequency 0.5), and evaluates $C_p$ over the airfoil suction side for simulation \#5 with reduced frequency 0.5.}
    \label{fig:reduced_frequency}
\end{figure}

To illustrate how divergence of marginal label distribution undermines generalizability, even in the presence of domain-invariant features, we base our discussion on airfoil kinematics variation. We begin by considering two different models, one trained on \circleBlue \space DS1 dataset and the other trained on \circleCyan \space DS2. Both datasets contain simulations of an SD7003 airfoil performing periodic and ramp-type motions, however, in \circleCyan \space DS2 we add an extra simulation with reduced frequency $0.5$
\footnote{Reduced frequency is a parameter used for the periodic motion and is defined as $k = \omega H / (2 U_{\infty})$, where $\omega$ is the circular frequency, $H$ is the amplitude of the plunge (we used $H = 0.5c$ in all periodic simulations), and $U_{\infty}$ is the freestream velocity. This is different from the parameter used for ramp motions. In this case, we use the pitch rate (indicated in Table \ref{table:simulations}), which is defined as $\Omega^{+} = \omega c / U_{\infty}$. Here, $\omega$ is the angular velocity of the airfoil and $c$ is the airfoil chord.}.
Notice that the only reduced frequency present in \circleBlue \space DS1 is $0.25$. Then, in Fig. \ref{fig:reduced_frequency} we operate these systems on a simulation with reduced frequency $0.5$, but at a different Mach number from that used in \circleCyan \space DS2. At such higher reduced frequency, the flow separates over the pressure side of the airfoil, as observed in the leftmost plot of the figure. The fact that the separation is not symmetrical is due to the airfoil having a static angle of attack of 8 deg. in our simulations\footnote{The reason for the dataset being formed by simulations with this angle of attack is that they were compared with the literature in other works by the authors.}.

When the flow separates on the airfoil pressure side, the suction is diverted to this side causing the pressure coefficient to become positive near the top of the leading edge (see the leftmost plot in Fig. \ref{fig:reduced_frequency}). Two problems occur here: (1) The color range for the pressure coefficient has its maximum set at 0, which leads to a concept shift due to color saturation; and (2) there is a misalignment of label distribution between source and target domains. Precisely, the label shift has a particular form of high-dimensional interval shift. This means that, although this simulation presents flow structures similar to those of the training set (domain-invariant features), the network trained on \circleBlue \space DS1 never experienced a positive value of $C_p$ on the airfoil suction side, as seen in the leftmost plot. In other words, many probes placed approximately in between $x/c = [0, \ 0.4]$ experience an interval shift. As a consequence, the divergence in the label distribution causes the model to fail to accurately predict the forces in the region close to the leading edge. Such misalignment is also observed in the second plot of Fig. \ref{fig:reduced_frequency}. In this case, the network never experienced a large and strong vortex structure so early in the dynamic stall development, as occurs with increasing reduced frequency. Despite this, the model trained on \circleBlue \space DS1 still predicts loading very well. But, after including one case of such variability in the dataset (\circleCyan \space DS2) the generalization improved. Finally, in the rightmost plot, both models perform similarly since this image is aligned with the source distribution.

It is important to highlight that, even in a flow with kinematic or fluid parameters very different from those with which the network was trained, there may be an intersection in the distribution of the flow features and their respective induced aerodynamic responses where the prediction will be accurate. To get an idea of the moment in which such an intersection occurs, let us take as an example the result above of the networks trained with \circleBlue \space DS1 and \circleCyan \space DS2 running the inference in simulation \#5, but now with a different objective in mind: to predict the aerodynamic coefficients. Notice that all the results we have shown so far are for the $C_p$ distribution over the suction side of the airfoil. The reason for building a surrogate model capable of estimating this quantity revolves around the complexity and/or cost of acquiring it in experimental settings. Nevertheless, once the network is capable of extracting features from flow visualizations, its dense layers can be designed for other tasks \citep{oquab2014learning}, such as predicting aerodynamic coefficients (lift, drag and quarter-chord pitch moment -- $C_l$, $C_d$ and $C_m$, respectively). This is done by changing the number of output layer nodes to 3 (one for each coefficient) and retraining the model.

\begin{figure}[!htb]
    \centering
    \includegraphics[width=\textwidth]{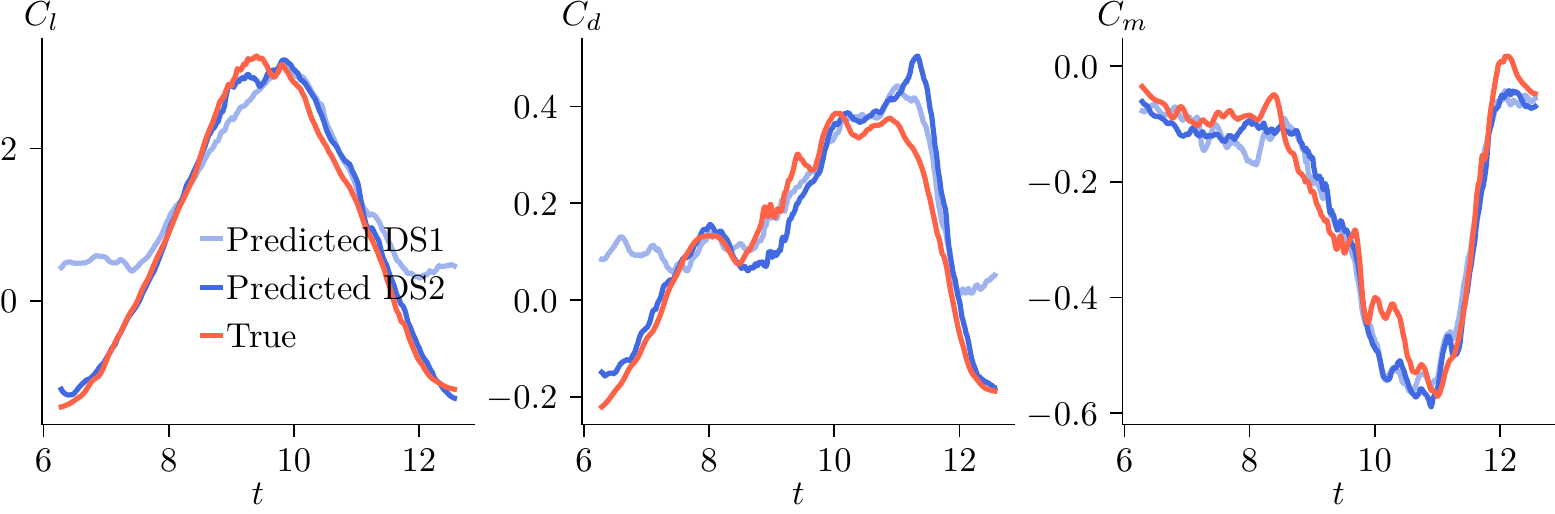}
    \caption{Effect of varying reduced frequency. The model is trained on \circleBlue \space DS1 (reduced frequency 0.25) or \circleCyan \space DS2 (adding a simulation with reduced frequency 0.5), and evaluates lift, drag and pitch moment coefficient for simulation \#5 with reduced frequency 0.5.}
    \label{fig:reduced_frequency_coeffs}
\end{figure}

The result is shown in Fig. \ref{fig:reduced_frequency_coeffs}, from which we see that the model trained on \circleBlue \space DS1 is not able to reconcile the lift, drag and pitch moment coefficients for the entire motion history for a case of a higher reduced frequency (simulation \# 5 was never seen by the network). However, in the interval from $8 \lessapprox t \lessapprox  11$, the source and target distributions are better aligned and the predictions are in good agreement. In fact, this model never experienced a situation where very low or negative lift and drag occur. These values come from the low pressure core that is formed on the pressure side of the airfoil that moves at this high reduced frequency, as mentioned earlier. Because it is a very different flow condition from those present in \circleBlue \space DS1, the network fails to predict the motion periods in which the flow dynamics shifts from the suction side of the airfoil to the pressure side. After adding one example of this semantic nuisance (\circleCyan \space DS2), the agreement between the predicted and true labels in Fig. \ref{fig:reduced_frequency_coeffs} improves considerably, especially for the pitch and drag coefficients. At this point, it is interesting to note that both the interpolation and extrapolation of the source domain lead to meaningful errors in the predicted drag, although the main trend is preserved. This error is probably related to the role of viscous effects in the calculation of the drag force, where it is more prominent.

To reinforce our arguments about including sources of variability, let us consider other variations of airfoil motion in the construction of the dataset. Here, we pose the following question: How is the inference of the network on airfoils in ramp-like motions if we train it only using periodic motions (\circleGray \space DS4), and vice-versa (\circleBlack \space DS5)? The flowfields are fundamentally different between these two types of motions. For ramp motions, the airfoil always presents a separated flow along the airfoil suction side, even before starting its movement. Furthermore, at high effective angles of attack, 
a strong trailing-edge vortex is formed \citep{miottoAIAAJ2022}. For the periodic simulations, in turn, the airfoil movement induces flow reattachment and the trailing-edge vortex is much weaker \citep{miottoJFM2022}. So, although in both cases there are common structures in the flowfield, such as the dynamic stall vortex and the trailing-edge vortex, the underlying aerodynamic response (marginal label distribution) can be substantially different for some samples taken from them. 

\begin{figure}[!htb]
    \centering
    \includegraphics[width=\textwidth]{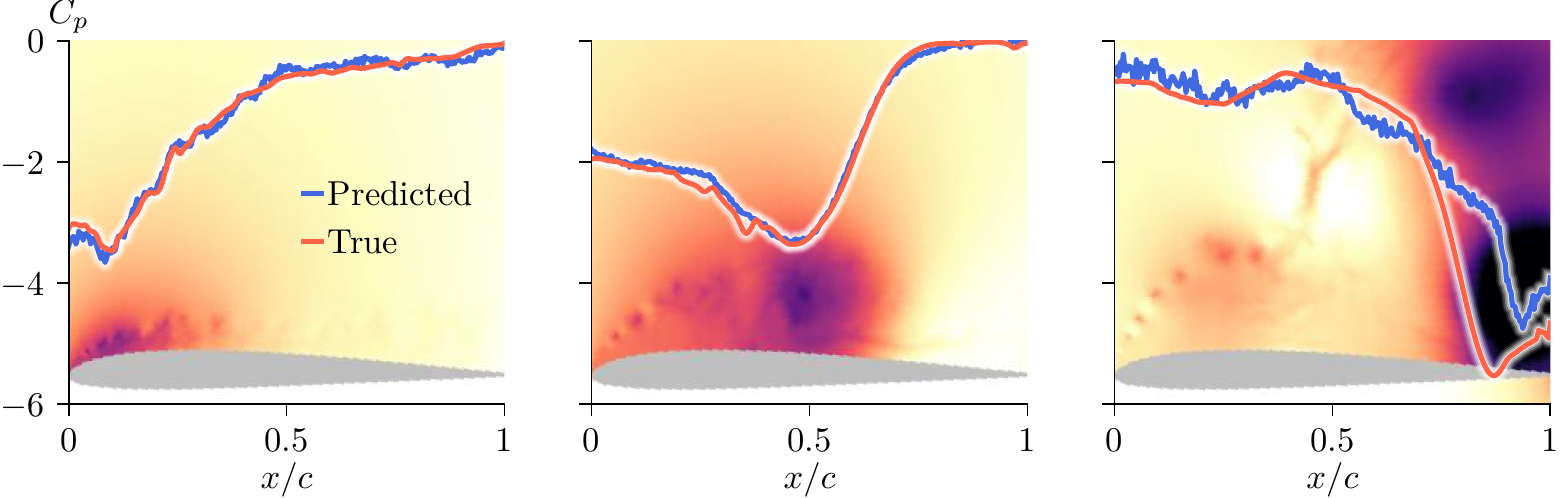}
    \caption{Effect of varying motion. The model is trained on \circleGray \space DS4 (periodic motion), and evaluates the $C_p$ distribution for simulation \#6, which consists of a ramp motion.}
    \label{fig:Cp_distribution_ramp_trained_on_periodic}
\end{figure}

Figure \ref{fig:Cp_distribution_ramp_trained_on_periodic} shows results for the neural network trained only with periodic data (\circleGray \space DS4) running inference on simulation \#6, which consists of a ramp movement. The first two plots of the figure demonstrate that the emergence and passage of the dynamic stall vortex under the airfoil are very well captured by the neural network. However, the model fails to estimate the airfoil loading from the moment the trailing edge vortex develops, as shown in the rightmost plot. Here, both the misalignment between the source and target domains and the trailing-edge vortex saturation problem impair generalization. Despite this, the pressure distribution trend is still maintained.

\begin{figure}[!htb]
    \centering
    \includegraphics[width=\textwidth]{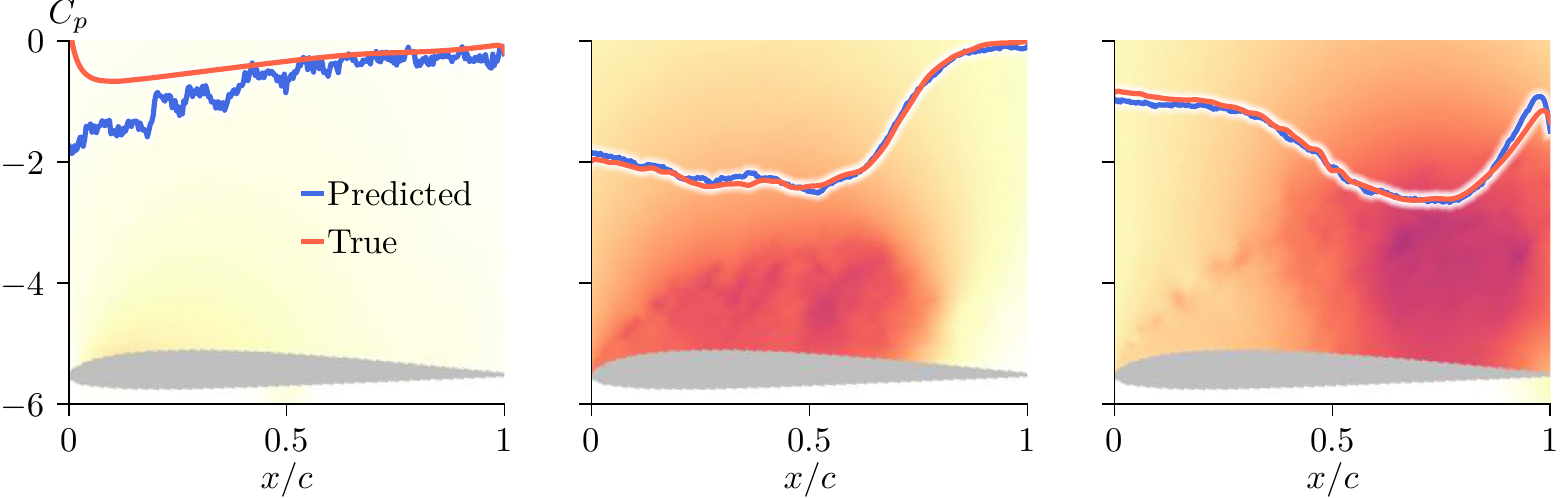}
    \caption{Effect of varying motion. The model is trained on \circleBlack \space DS5 (ramp motion), and evaluates the $C_p$ distribution for simulation \#3, which consists of a periodic motion.}
    \label{fig:Cp_distribution_periodic_trained_on_ramp}
\end{figure}

A similar conclusion is obtained when the model trained only with ramp data (\circleBlack \space DS5) operates in a periodic case. In this sense, Fig. \ref{fig:Cp_distribution_periodic_trained_on_ramp} shows the results for snapshots of simulation \#3, through which the two plots on the right show that the passage of the dynamic stall vortex is well represented by the network. However, when the flow is attached (the leftmost plot), the network prediction fails due to presence of both a covariate shift and an interval shift. As we observed, these shifts often arise in practical regression settings that need moderate extrapolation and interpolation.

\subsection{Synthesizing fluid properties}
\label{sec:synthetic_fluid_properties}

In Sec. \ref{sec:different_property_as_input}, we showed that the pressure coefficient field is preferable to velocity when it comes to building our regression model. But the role that pressure plays in flow analyses goes beyond simply obtaining aerodynamic loadings. In this session we will show how neural networks can be used to extract the pressure field from the fluid flow velocities. This is important because, while this variable is readily available in numerical simulations, its determination in a experimental setting is more challenging, particularly in compressible flows, and requires appropriate techniques \citep{VanderKindere2019}. Thus, in this section, we present an alternative to determine the pressure field from velocity data, complementing the previous regression model.

The approach works similarly to what we have already discussed, in the sense that the available data (whether from numerical simulations or from experiments) can be used to train the model for later use in experiments. For example, if provided with numerical data where the thermodynamic properties can be easily obtained, these data can be used to train the regressor. In the case of having only PIV data, the pressure would first need to be determined using the techniques described by \citet{VanderKindere2019} to then train the model. In the present work, we demonstrate the application of this technique only in the numerical scope, since these are the data available. But, as \citet{Visbal_2011} shows, PIV and numerical visualizations are comparable to each other and, therefore, this does not change our conclusions. In addition, if experimentally acquired flowfields are corrupted with incorrect and missing entries, the recent technique by \citet{denoise_PIV} can be used to improve the flowfield data quality.

\begin{figure}[!htb]
    \centering
    \includegraphics[width=\textwidth]{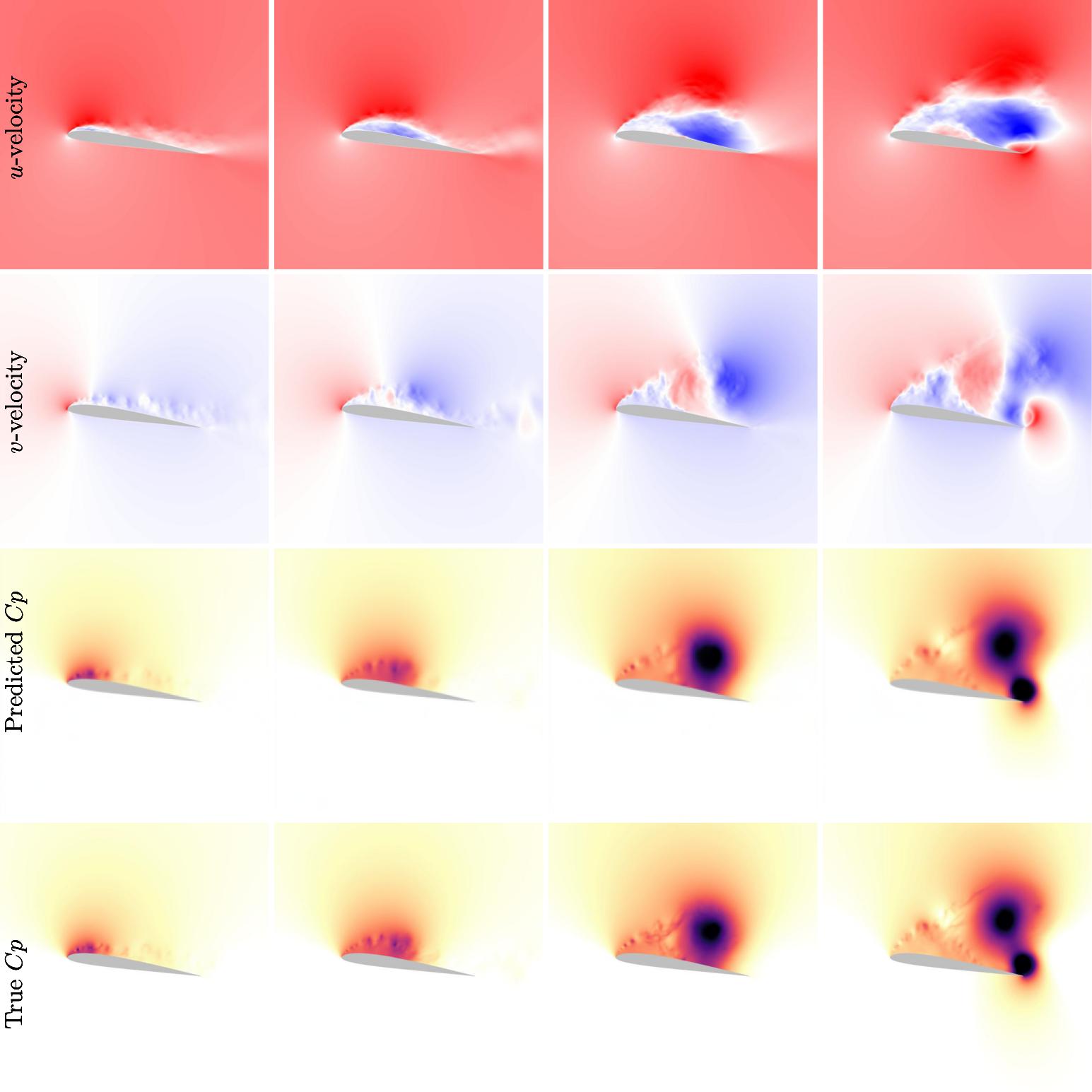}
    \caption{Output images from model trained on \circleBlue \space DS1 ($M_{\infty} = 0.1$ and $0.4$) using $u$- and $v$-velocity components as input. The model evaluates the $C_p$ flowfield for arbitrary snapshots of simulation \#10 with $M_{\infty} = 0.2$.}
    \label{fig:unet_predictions}
\end{figure}

Here, we draw on image-to-image translation to amplify the scope of information extracted in experimental fluid mechanics. Particularly, given the source domain of images of the velocity field concatenated channelwise, we train a U-Net \citep{Ronneberger_etal_2015} to generate synthetic pressure coefficient fields. For this, the same dataset \circleBlue \space DS1 from the regression model was used, and after 84 epochs, we reached a validation mean squared error of approximately $1.35\mathrm{e}-5$, which we consider satisfactory. This training took about 4.5 hours in a NVIDIA Tesla A100. Some results are presented in Fig. \ref{fig:unet_predictions}, where we show the input velocities, the actual pressure coefficient fields and those predicted by the U-Net. These results are displayed columnwise, each column corresponding to a snapshot randomly selected from simulation \#10 with $M_{\infty} = 0.2$. Therefore, we are interpolating between different Mach numbers. 

\begin{figure}[!htb]
    \centering
    \includegraphics[width=\textwidth]{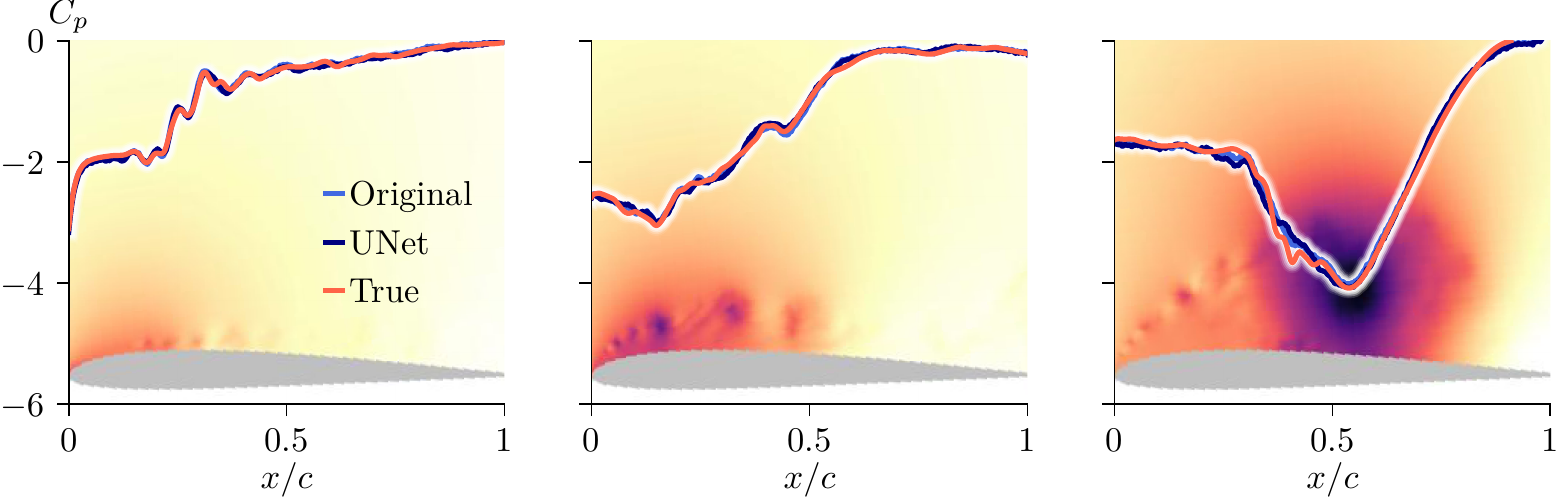}
    \caption{Calculation of surface $C_p$ distribution from output images. The model is trained on \circleBlue \space DS1 ($M_{\infty} = 0.1$ and $0.4$) using the artificially generated $C_p$ field as input, and evaluates the $C_p$ distribution for arbitrary snapshots for simulation \#10 with $M_{\infty} = 0.2$.}
    \label{fig:unet_coeffs}
\end{figure}

An excellent agreement is observed between the true and predicted fields in Fig. \ref{fig:unet_predictions}, demonstrating that a simple encoder-decoder architecture can correctly map the input attribute values to the corresponding synthetic image. However, some fine details are missing out from the predictions. 
Nevertheless, these synthetic fields provide good results when used as input to the regression model built previously. To illustrate this, in Fig. \ref{fig:unet_coeffs} we evaluate the accuracy of using the synthetic $C_p$ images of some snapshots from simulation \#10 and also compared them with the results obtained using the real pressure field as input and  with the true value of the $C_p$ distribution over the airfoil suction side. Clearly, the synthetic field produces an almost identical estimate of the pressure distribution when feeding the true pressure field into the network. Thus, we demonstrate that image-to-image translation (U-Net) and regression (ViT-B/16) networks can be combined to improve airfoil loading prediction from velocity data. We believe that this framework can improve the arsenal of tools available for analyzing unsteady flows with a broad range of spatial and temporal scales.

\section{Conclusions}



Based on unsteady flowfields, a pre-trained ViT-B/16 network renders the backbone of a neural network model that links the existing flow structures to the aerodynamic response of the airfoil. Here, we focus on the $C_p$ surface distribution, but we also show that it is possible to obtain other quantities such as aerodynamic coefficients just by retraining the network with a different number of outputs in the network. The ViT correctly inferes the attributes present in the flow image even in a compressible flow regime for which no annotations are given. As a result, an excellent agreement between predicted and ground truth values is obtained for the $C_p$ over the airfoil suction side. Similar conclusions are reached with CNNs (not shown here for brevity), but ViT outperforms them in the regression task. This fact demonstrates that ViT- or CNN-based models can be used to interpolate between flow parameters. The ability of the neural network to extrapolate the source domain is also investigated by varying Mach and Reynolds numbers, besides the airfoil kinematics. In this sense, it is demonstrated that using pressure as input to the network instead of the velocity field improves generalization. In principle, this makes sense, since pressure is directly related to forces.

When encountering out-of-distribution data ({\em i.e.}, when the target flowfield contains never seen semantics), the model is prone to error. So, in this work, we aim to establish to what degree it impacts the model to take steps to resolve the issues and improve the model’s accuracy. By training our network on different types of simulation parameters, we demonstrate that the position and morphology of the dynamic stall vortex with respect to the airfoil, as well as the trailing edge vortex, are sources of variability that must be learned for the model to better generalize. This ensures that marginal label distribution is aligned across domains, which is known to be a necessary condition for generalization when learning domain-invariant representations. Fortunately, though, in fluid mechanics most of the nuisance variability is already known and can be dealt with at the outset.

In addition to providing evidence that not including sources of variability in the training set can lead to a deterioration in model performance, we purposely subject it to other sources of error to assess its robustness. For example, the saturation color range of the images is maintained and the airfoil is not perfectly framed in the image, in the sense that there is some padding extending from the airfoil limits. These conditions are likely to occur in practice and, therefore, must be taken into account when judging the result of inferences. In this sense, we show that the presence of saturated regions near the airfoil surface leads to a concept shift, which is when the same input results in different outputs. Thus, we can say that the saturated regions are ambiguous representations. The non-framing of the airfoil in the image, in turn, results in a low spatial resolution that causes the smoothing of fine-grained scale phenomena, which are more likely to exist at high Reynolds numbers. These observations can be taken as good practices when training the neural network. But even in the presence of these errors, the network inferences show good agreement with the true results.

Finally, we also demonstrate that the velocity can be used to synthesize any corresponding physical quantity through an image-to-image translation approach. Here, the mapping between velocity and pressure coefficient field is used as example for it is preferable to velocity when it comes to building our regression model. The results using the synthetic image show an excellent agreement with those obtained from the actual pressure field, proving to be an interesting pre-processing technique for regression.



\section*{Acknowledgments}

The authors acknowledge Fundação de Amparo à Pesquisa do Estado de São Paulo, FAPESP, for supporting the present work under research grants No. 2013/08293-7, 2013/07375-0 and 2021/06448-0. FAPESP is also acknowledged for the fellowship provided to the first author under grant 2022/09196-4. The CEPID-CeMEAI cluster Euler, and cluster SDumont are acknowledged for providing the computational resources for this research. Finally, Conselho Nacional de Desenvolvimento Científico e Tecnológico, CNPq, is also acknowledged for supporting this research under Grants No. 407842/2018-7 and 308017/2021-8.

\bibliographystyle{elsarticle-num-names} 
\bibliography{sample}

\end{document}